\documentclass[aps,prd,a4paper,onecolumn,amsmath,showpacs,superscriptaddress,nofootinbib,preprintnumbers,notitlepage]{revtex4-1}
\usepackage{amssymb,amsmath}
\usepackage{graphicx}
\usepackage{color}
\usepackage{dcolumn}
\usepackage{pgfplots}
\usepackage{booktabs}
\usepackage{multirow}
\usepackage{dcolumn}
\usepackage{amsmath}
\usepackage{mathtools}
\usepackage{amsfonts}
\usepackage{amssymb}
\usepackage{epstopdf}
\usepackage{bm}
\usepackage{siunitx}
\usepackage{braket}
\usepackage{enumitem}
\usepackage{soul}
\usepackage{ulem}
\usepackage{color}
\usepackage{transparent}
\usepackage{pifont}
\usepackage[font={small}]{caption}
\newcommand\rf[1]{(\ref{eq:#1})}
\newcommand\lab[1]{\label{eq:#1}}

\newcommand\br{\begin{eqnarray}}
\newcommand\er{\end{eqnarray}}
\newcommand\be{\begin{equation}}
\newcommand\ee{\end{equation}}

\newcommand\lb{\lbrack}
\newcommand\rb{\rbrack}

\newcommand\bc{\begin{center}}
\newcommand\ec{\end{center}}

\newcommand\partder[2]{\frac{{\partial {#1}}}{{\partial {#2}}}}

\newcommand\eps{\epsilon}
\newcommand\vareps{\varepsilon}

\newcommand\G{\Gamma}

\renewcommand\k{\kappa}
\renewcommand\l{\lambda}

\newcommand\m{\mu}
\newcommand\n{\nu}

\renewcommand\P{\Phi}
\newcommand\pa{\partial}

\newcommand{\ct}[1]{\cite{#1}}
\newcommand{\bib}[1]{\bibitem{#1}}

\newcommand\PRD[3]{\textsl{Phys. Rev.} \textbf{D#1} (#2) #3}

\newcommand\PLB[3]{\textsl{Phys. Lett.} \textbf{#1B} (#2) #3}
\newcommand\CQG[3]{\textsl{Class. Quantum Grav.} \textbf{#1} (#2) #3}

\newcommand\IJMPD[3]{\textsl{Int. J. Mod. Phys.} \textbf{D#1} (#2) #3}

\newcommand\MPLA[3]{\textsl{Mod. Phys. Lett.} \textbf{A#1} (#2) #3}

\begin{document}
\title {Signed  Coordinate Invariance, invariant lagrangians and manifolds, the time problem in quantum cosmology, quantum space time,  spacetimes and antispacetimes.
}

\author{Eduardo Guendelman}
\email{guendel@bgu.ac.il}
\affiliation{Department of Physics, Ben-Gurion University of the Negev, Beer-Sheva, Israel.\\}
\affiliation{Frankfurt Institute for Advanced Studies (FIAS),
Ruth-Moufang-Strasse 1, 60438 Frankfurt am Main, Germany.\\}
\affiliation{Bahamas Advanced Study Institute and Conferences, 
4A Ocean Heights, Hill View Circle, Stella Maris, Long Island, The Bahamas.
}
\begin{abstract}
Standard general coordinate invariance for the volume element is extended to general coordinate transformations that have a negative jacobian. This is possible by introducing a non Riemannian Measure of integration, which transforms according to the jacobian of the coordinate transformation, not the absolute value of the  jacobian of the coordinate transformation as it  is the case with $\sqrt{-g}$. For some signed general transformations a change of boundary conditions is involved in GR and restoring the original limits of integrations can restore symmetry of the action. The discussion totally changes when a non Riemannian Measure of integration is introduced, in the case of signed general coordinate transformations,  and even when  we have, boundaries,  since the modified measure is constructed out of scalar fields becomes also the integration manifold, Since the measure fields that define the non metric measure are scalars.  if a signed general coordinate transformations is considered, there is no change in the measure fields, indicating strict invariance of both lagrangian density and integration manifold, implying boundary terms for manifolds with boundaries in coordinate space to be irrelevant . This analysis can be applied to give a framework where in certain scenarios, like t can give something similar to the  Linde´s  Universe multiplication from first principles,  Linde  assumes that together with the observed universe there is an additional one with analogous matter and gravity content but with opposite action. The Linde model also has a (restricted) version of the signed general coordinate invariance (without affecting Boundary Conditions) . We  consistently formulate the non Riemannian measure theory extension of General Relativity, that could be related to Linde´s model,  although there are some fundamental differences with Linde´s scenario. like there is here  a local formulation. The assumption of a coordinate independent  measure of integration appears also useful in the study of Baby Universe Creation. The formulation provides also a new natural way to address the problem of time in Quantum Cosmology,  a  proposal for a quantum space time and identification of space times and anti space times states in the gravitational theory, which in fact relates very much to the model of Linde, which can be interpreted now as a non local coexistence of space time  and anti spacetime.
\end{abstract}
\maketitle
\section{Introduction}
\label{intro}

In an interesting paper, Linde formulated a model that claims to resolve the cosmological constant problem \cite{Lindemultiplication}.  This requires the existence of two universes,
each with its own set of coordinates $x^\mu$ and $y^\mu$ containing matter and gravity components that mirror each other, but with the corresponding actions having opposite signs, like spacetime and an antispacetime, 
as in 

\be \lab{lindedoubleaction}
S = \int d^4 x d^4 y {\sqrt{-g(x)}} {\sqrt{-\bar{g}(y)}} (\frac{M_P^2}{16\pi} R(x) + L(\phi(x)) - \frac{M_P^2}{16\pi} R(y) - L((\bar{\phi}(y))
\ee
$R(x)$ is defined in terms of a $g_{\m\n}$ metric, while $ R(y)$  is defined in terms of a  $\bar{g}_{\m\n}$ metric and  
where $L(x)$ and $L(y)$ have exactly the same functional form with respect to their corresponding mirror fields, like for a scalar field $\phi(x)$, there will be a potential  $V(\phi(x))$, same with kinetic terms, etc. that are 
appearing in  $L(x)$ , while in $L(y)$  there will be corresponding field $\bar{\phi}(y)$ with a potential $V(\bar{\phi}(y))$, then in $L(y)$ the metric $\bar{g}_{\m\n}$ appears instead of 
the metric $g_{\m\n}$, then  the theory is obviously invariant under $V \rightarrow V + constant $ \cite{Lindemultiplication}. This non local coexistece of a spacetime and an antispacetime was shown by Linde to have remarkable properies concerning its behavior with respect to the cosmological constant problem. As we will see , we can think of this as  having regions where the measure of integration
can change sign, an effect that must take place at the same time as we double the space time, to realize  Linde´s ideas. The model by Linde is however non local, so this is an aspect that is not desirable, we will explore how models inspired by modified measures (non Riemannian) avoids this.

Another area where our formalism can help is the more complicated process of quantum creation of inflationary bubbles. In the most popular paradigm  for the early universe, it is postulated that 
the universe suffers a period of exponential expansion called ``inflation'' introduced by Guth \cite{Guth} , Starobinsky \cite {starobinsky} , Linde \cite {Linde} , \cite{ALBRECHTANDSTEINHARDT},  (see also the books \ct{early-univ,primordial} and references therein).
The possibility of a "local" version of inflation, that is an inflationary bubble, was studied in \cite{BLAUGUENDELMANGUTH} a process which was later generalized so as to include the quantum creation of such bubbles in \cite{FARHIGUVEN}, for an alternative treatment of creation of baby universes see \cite{FischlerPolchinski}. For a popular review on the subject of Baby Universes and further references see \cite{ZEEYA} .
An outstanding feature of \cite{FARHIGUVEN} is that in their euclidean tunneling solution there are regions where the measure defining the volume element must be negative. However this seems hard to understand standard Riemannian measure $\sqrt{-g}$
cannot behave in this way, nevertheless \cite{FARHIGUVEN} gives very good reasons that this should be the correct behavior of the measure in the euclidean tunneling solution. For other approaches to the quantum creation of a baby universe see \cite{FischlerPolchinski}.

Here we will address Linde´s scenario and tentativelly also the much more complicated baby universe creation in the framework of the metric independent non Riemannian measures, which has been used  for the construction of modified gravity theories
Refs.\ct{TMT-orig-1}-\ct{TMT-orig-3} (see also 
Refs.\ct{TMT-recent-1-a}-\ct{TMT-recent-2}), in some instances we have included the standard measure as well, where the standard Riemannian integration measure  might also contain a Weyl-scale symmetry preserving $R^2$-term \ct{TMT-orig-3}. Some applications have been, (i) $D=4$-dimensional models of gravity and matter fields containing  the new measure of integration appear to be promising candidates for resolution  of the dark energy and dark matter problems, the fifth force problem, and a natural mechanism for spontaneous breakdown of global Weyl-scale symmetry \ct{TMT-orig-1}-\ct{TMT-recent-2}, (ii) To study in Ref.\ct{susy-break} of modified supergravity models with an alternative non-Riemannian volume form on the space-time manifold, non singular emergent models leading to inflation and then decaying into a dark enery and dark matter phase  \cite{ourquintessence} - \cite{ourquintessencewithEDE},(iv) 
Gravity-Assisted Emergent Higgs Mechanism in the Post-Inflationary Epoch, \cite{Gravityassisted},  (v) To study of reparametrization invariant theories of extended objects (strings and branes) based on employing of a modified non-Riemannian 
world-sheet/world-volume integration measure \ct{mstring}, \ct{nishino-rajpoot}, leads to dynamically 
induced variable string/brane tension and to string models of non-abelian 
confinement, interesting consequences from the modified measures spectrum \ct{mstringspectrum}, and construction of new braneworld scenarios \ct{mstringbranes},   .
Modifed Measures Theories have been  discussed as effective theories for causal fermion theories \ct{MMT}. 

We will see here how in this framework one can construct a General coordinate invariant which has extended general coordinate transformations that includes also transformations with a negative jacobian, this is possible by introducing a non Riemannian Measure of integration, which transforms according to the jacobian of the coordinate transformation, not the absolute value of the  jacobian of the coordinate transformation as it  is the case with $\sqrt{-g}$. This analysis can be applied to give a framework for the Farhi, Guven and Guth treatment of the Quantum creation of a baby Universe, since here the new theory certainly allows for negative measures of integration required by Linde´s universe multiplication and the Farhi, Guth, Guven tunneling solutions where the measure of integration also can be negative.
\section{ General Relativity and other  theories use a  Riemannian volume element that is not invariant under signed general coordinate transformations}
\label{GR}
The action of GR, and other theories that use the standard Riemannian volume element  $ d^4 x {\sqrt{-g}}$  is of the form,
\be
S = \int d^4 x {\sqrt{-g}} L
\lab{GRL}
\ee

where $L$ is a generally coordinate invariant lagrangian.
Now notice that under a general coordinate transformation, 
$$ d^4 x \rightarrow Jd^4 x $$ , while 
$$ \sqrt{-g} \rightarrow  \mid J \mid ^{-1}\sqrt{-g}$$
where $J$ is the jacobian of the transformation and $ \mid J \mid$ is the absolute value of the transformation. Therefore $d^4 x {\sqrt{-g}} \rightarrow  \frac{J}{ \mid J \mid} d^4 x {\sqrt{-g}} $, so invariance is achieved only for $J = \mid J \mid$, that is if $J>0$, that is signed general coordinate transformations are excluded.

One could argue that when taking the square root of the determinant of the metric one may choose the negative solution when it suit us, but this would be an arbitrary procedure if no specific rule is given to choose the positive or the negative root. We choose instead to declare that $ \sqrt{-g} $ is always positive and replace it in the measure by something else whose sign is well defined. 
\subsection{Invariance of the action with non invariant lagrangian density  (integrand) and compensating non invariant manifold of integration?}

If conditions are optimal, the non invariance of the lagrangian density  (integrand) , which includes the measure, in a signed coordinate transformation, could be compensated by the non invariance of the manifold of integration. For example, in a time reversal transformation, the integrand will change sign, but , if there no obstructions, and we can then change the limits of integrations, the exchange of the limits of integrations will involve an additional exchange of signs that can compensate for the sign change in the integrand.

If the manifold has boundaries in coordinate space, introducing coordinate transformations that change the boundaries involve compensationg terms at the boundaries, even when the transformations are infinitesimal,
so at this point the invariance of the action becomes complicated and problematic, in particular in cosmology where the universe may have a beginning in time.

As we discuss in the next sections, where we will discuss a possible realization of the  invariance under signed general coordinate transformations in the context of the non local Linde Universe multiplication model, without invoking a change in the manifold of integration, but rather transforming the field variables,  and then in the following section a return to a local theory by the use of the modified measure formalism.

\section{Invariance of signed general coordinate transformations in  the non local Linde Universe multiplication model, or spacetime antispace model }

The Linde non local model can offer a limited way out to obtain signed general coordinate invariance, so we can allow general coordinate transformations where the jacobian of the transformation, say in the $x$ space , is negative, but still we do not consider the possibility that it could change from positive to negative.





As long as the jacobian is uniformly negative over all space, invariance will be achieved if the same transformation is performed both in the $x$ and $y$ coordinates
In these cases, we do not invoke any transformation or change in the manifold of integration, which we have argued is a questionable operation.

As we will see in the next section, the use of Metric Independent Non-Riemannian Volume-Forms and Volume elements allows us to resolve this issue with no such restrictions on possible changes of signs of the jacobian of the coordinate transformation in different regions of space time and without invoking non local actions.

\section{Metric Independent Non-Riemannian Volume-Forms and Volume elements invariant under signed general coordinate transformations}
One can define a metric independent measure from a totally anti symmetric tensor gauge field, for example
\be
 \Phi (A) = \frac{1}{3!}\vareps^{\m\n\k\l} \pa_\m A_{\n\k\l} \quad ,
\lab{Phi}
\ee

Then, under a general coordinate transformation $$\Phi (A) \rightarrow J^{-1}\Phi (A) $$. . Therefore $d^4 x \Phi(A)  \rightarrow  d^4 x \Phi(A) $, so invariance is achieved regardless of the sign of $J$.

\section{Theory using Metric Independent Non-Riemannian Volume-Forms}
\label{TMMT}

First we review our previous papers where we have considered  the  action 
of the general form involving two independent non-metric integration
measure densities generalizing the model analyzed in \ct{quintess} is given by 
\be
S = \int d^4 x\,\P_1 (A) \Bigl\lb R + L^{(1)} \Bigr\rb +  
\int d^4 x\,\P_2 (B) \Bigl\lb L^{(2)} + \eps R^2 + 
\frac{\P (H)}{\sqrt{-g}}\Bigr\rb \; .
\lab{TMMT1}
\ee

Here the following definitions  are used:

\begin{itemize}
\item
The quantities $\P_{1}(A)$ and $\P_2 (B)$ are two densities and these are  independent non-metric volume-forms defined in terms of field-strengths of two auxiliary 3-index antisymmetric
tensor gauge fields
\be
\P_1 (A) = \frac{1}{3!}\vareps^{\m\n\k\l} \pa_\m A_{\n\k\l} \quad ,\quad
\P_2 (B) = \frac{1}{3!}\vareps^{\m\n\k\l} \pa_\m B_{\n\k\l} \; .
\lab{Phi-1-2}
\ee
The density $\P (H)$ denotes  the dual field strength of a third auxiliary 3-index antisymmetric
tensor 
\be
\P (H) = \frac{1}{3!}\vareps^{\m\n\k\l} \pa_\m H_{\n\k\l} \; .
\lab{Phi-H}
\ee

\item
The scalar curvature $R = g^{\m\n} R_{\m\n}(\G)$ and the Ricci tensor $R_{\m\n}(\G)$ are defined in the first-order (Palatini) formalism, in which the affine
connection $\G^\m_{\n\l}$ is \textsl{a priori} independent of the metric $g_{\m\n}$.
 Let us recall that $R+R^2$ gravity within the
second order formalism  was originally
developed in \ct{starobinsky}.
\item
The two different Lagrangians $L^{(1,2)}$ correspond to two  matter field Lagrangians 
\end{itemize}
On the other hand, the variation of  \rf{TMMT1} w.r.t. auxiliary tensors 
$A_{\m\n\l}$, $B_{\m\n\l}$ and $H_{\m\n\l}$ becomes
\be
\pa_\m \Bigl\lb R + L^{(1)} \Bigr\rb = 0 \quad, \quad
\pa_\m \Bigl\lb L^{(2)} + \eps R^2 + \frac{\P (H)}{\sqrt{-g}}\Bigr\rb = 0 
\quad, \quad \pa_\m \Bigl(\frac{\P_2 (B)}{\sqrt{-g}}\Bigr) = 0 \; ,
\lab{A-B-H-eqs}
\ee
whose solutions are
\be
\frac{\P_2 (B)}{\sqrt{-g}} \equiv \chi_2 = {\rm const} \;\; ,\;\;
R + L^{(1)} = - M_1 = {\rm const} \;\; ,\;\; 
L^{(2)} + \eps R^2 + \frac{\P (H)}{\sqrt{-g}} = - M_2  = {\rm const} \; .
\lab{integr-const1}
\ee
Here the parameters $M_1$ and $M_2$ are arbitrary dimensionful and the quantity $\chi_2$ corresponds to an
arbitrary dimensionless integration constant. 

The resulting theory is called a Two Measure Theory, due to the presence of the Two measures  $\P_1 (A) $ and $\P_2 (A) $. But for the purpose of this paper this is two general, since we want to restrict to a theory that will give us ordinary General Relativity, and we want to keep the general coordinate invariance under signed general coordinate invariance. 

For obtaining GR dynamics, we can restrict to one measure, so let us take 

$$ \P_1 (A)= \P_2 (B) = \Omega $$

also to make some contact for example with  \cite{FARHIGUVEN} , where an additional set of four fields is introduced,  we express $ \Phi $ in terms of
four scalar fields
\be
\Omega = \frac{1}{3!}\vareps^{\m\n\k\l}\vareps^{abcd} \pa_\m\varphi_a \pa_\n\varphi_b \pa_\k\varphi_c \pa_\l\varphi_d \quad  
\lab{omega}
\ee
(one has to point out that in the earlier formulations of modified measures theories we used the 4 scalar field representation for the measure, see \cite{TMT-orig-1} , )
The mapping of the four scalars to the coordinates $x^\mu$ may be topologically non trivial, as in \cite{FARHIGUVEN} and this multivaluedness could be of use to obtain Linde´s Universe multiplication as well. Finally, we have to correct the equation
\be
\frac{\P_2 (B)}{\sqrt{-g}} \equiv \chi_2 = {\rm const}  .
\lab{eq.tobe corrected}
\ee
for another equation the will be invariant under signed general coordinate invariant transformations, which will be 
\be
\frac{\Omega^2}{(-g)}\equiv \chi = K^2 =  {\rm const}  > 0  .
\lab{corrected}
\ee
without loss of generality we define $K$ to be positive.
The resulting action that replaces \rf{Phi-1-2} is, 
\be
S = \int d^4 x\,\Omega \Bigl\lb R + L \Bigr\rb +  
\int d^4 x\,\Omega^2 \Bigl\lb 
\frac{\P (H)}{{(-g)}}\Bigr\rb \; .
\lab{simpleGCISIGNED}
\ee
the density $\P (H)$ remains defined eq. \rf{Phi-H}
so the integration obtained from the variation of the $H$ gauge field is  eq. \rf{corrected} now.
The solution of  eq. \rf{corrected} are
\be
\frac{\Omega}{ \sqrt{(-g)}} = \pm{ K}   .
\lab{solutions of corrected}
\ee
where the sign in \rf{solutions of corrected} will  be dynamically determined, as opposed to the Linde scenario, where the negative measure, associated with the $y$ universe always exists. The non locality in the Linde´s  scenario could be replaced by the introduction of a new set of coordinates in a way such that
the $\varphi_a $ scalars  have a  relation to the coordinates $x$ to these scalars may be multi valued, What is no doubt is  similar is the existence of positive and negative measures of integration in both theories. In both cases it is related to signed  general coordinate transformation, although in the Linde model the sign has to be uniform over space time and in the modified measure case this restriction does not exist, in fact the sign of the jacobian of the transformation can vary from one region of space time to the other in the  non Riemannian measure formulation, 

Another possibility for a measure that would transform like the the jacobian of the coordinate transformation, not the absolute value of the jacobian,  would be the determinant of the vierbein. This will detroy however (up to a sign) the invariance of the theory under signed local Lorentz transformation of the vierbeins. that is Lorentz transformations with negative determinants, so, it is not a solution, rather we trade one asymmetry for another.

\section{The invariant scalar integration manifold and invariant lagrangian density}
\label{invariantscalarintegrationmanifold}

Notice that using the volume element converts the the integration over coordinates in the action into integration over scalar fields, since $$\Phi d^4x = d\varphi_1 d\varphi_2 d\varphi_3 d\varphi_4$$

The integration manifold existing in the four scalar field manifold is in fact completely unaffected by any coordinate transformation taking place in the $x$ space. The lagrangian density is also a scalar not affected by any coordinate transformation, the theory formulated in this way does not require any boundary terms if the boundaries are for example formulated in the scalar field space.

In the case of \rf{simpleGCISIGNED} for example, $$S= \int  d\varphi_1 d\varphi_2 d\varphi_3 d\varphi_4 L $$,
where
$$L=  \Bigl\lb R + L \Bigr\rb +  \,\Omega \Bigl\lb 
\frac{\P (H)}{{(-g)}}\Bigr\rb \; $$

\section{Gravitational Equations of motion}
\label{Einstein}

We start by considering the equation that results from the variation of the degrees of freedom that define the measure $\Omega$,  that is the scalar fields $\varphi_a$, these are,
\be
A^{\m}_a \pa_\m (R + L  +2 \Omega \frac{\P (H)}{{(-g)}}) = 0
\lab{EINSTEIN}
\ee
where 
\be
A^{\m a} =\frac{1}{3!}\vareps^{\m\n\k\l}\vareps^{abcd}  \pa_\n\varphi_b \pa_\k\varphi_c \pa_\l\varphi_d \quad  
\lab{AMATRIX}
\ee
Notice that the determinant of $A^{\m a}$ is proportional to $\Omega^3 $, so if the measure is not vanishing, the matrix $A^{\m a}$ is non singular and therefore $\pa_\m (R + L  +2 \Omega \frac{\P (H)}{{(-g)}}) = 0 $, 
so that,
\be
R + L  +2 \Omega \frac{\P (H)}{{(-g)}} = M = constant
\lab{M}
\ee

The variation with respect to the metric $g^{\m\n}$, we obtain.
\be
\Omega (R_{\m\n}+ \frac{\pa L}{\pa g^{\m\n} } )  +   g_{\m\n} \Omega^2 \frac{\P (H)}{{(-g)}} = 0
\lab{munu}
\ee
solving $\Omega \frac{\P (H)}{{(-g)}}$  from \rf{M} and inserting into \rf{munu}, we obtain,
\be
R_{\m\n} - \frac{1}{2}g_{\m\n} R  + \frac{1}{2} M g_{\m\n} + \frac{\pa L}{\pa g^{\m\n} } - \frac{1}{2}g_{\m\n} L = 0
\lab{EINSTEINLIKEEQ}
\ee
which gives exactly the form of Einstein equation with the canonical energy momentum defined from $L$

\be
T_{\m\n} = g_{\m\n} L - 2 \partder{}{g^{\m\n}} L \; .
\lab{EM-tensor}
\ee

The equations of motion of the connection (in the first order formalism) implies that the connection is the Levi Civita connection.  L can describe a scalar field with the potential and the term $\frac{1}{2} M $ can be interpreted as a shift of the scalar field potential by a constant or a floating contribution to the cosmological constant- In \cite{FARHIGUVEN}  the tunneling solution is described by considering another embedding space in addition to the standard $x^\m$ space,  the mapping between the two set of coordinates is multi valued , so that for each point in the  $x^\m$ space there are many points in the embedding space.
In the modified measure approach to this problem is best to consider the embedding space as the one defined by the four scalar fields $\varphi_a$ that define the measure $\Omega$ \rf{omega}. Therefore the most fundamental space is the  $\varphi_a$ , since only in this space we can formulate the full description and solution of the problem. The calculations in \cite{FARHIGUVEN} are consistent with GR,
just extended to allow for negative measures of integration, which is exactly what we have formulated here in the context of a modified measure theory.

One issue that should be addressed is that of the gauge fixing in the $\varphi_a$ space. Indeed, we notice that the only thing where these fields appear in the equations of motion is $\Omega$, but this quantity is invariant under volume preserving diffeomorphisms of the fields $\varphi_a$, $\varphi^\prime_a = \varphi^\prime_a(\varphi_a)$  which satisfy

\be \lab{VOLPRESDIFF}
\epsilon_{a_1 a_2 a_3 a_4}\frac{\partial{\varphi^\prime}_{b_1}}{\partial\varphi_{a_1}}\frac{\partial{\varphi^\prime}_{b_2}}{\partial\varphi_{a_2}}\frac{\partial{\varphi^\prime}_{b_3}}{\partial\varphi_{a_3}}\frac{\partial{\varphi^\prime}_{b_4}}{\partial\varphi_{a_4}} = \epsilon_{b_1 b_2 b_3 b_4}
\ee

so the study of the best gauge for the $\varphi_a$  fields for further comparison with the  $x^\m$ space could be a very important subject. Of course when we say that the mapping between the $\varphi_a$ and the $x^\m$ spaces, we want to exclude multi valuedness due to volume preserving diffeomorphisms of the fields $\varphi_a$, if for example different signs for $\Omega$ are associated to the same point in  $x^\m$ space, it is clear that there are at least two points  in $\varphi_a$ space associated to one point in  $x^\m$ space, and these two points in  the $\varphi_a$ are not related through a volume preserving diff. This could be an effect analogous to the Universe Multiplication of Linde. 
\section{Linde´s universe Multiplication and relation to a Brane Anti Brane system and measure field multivaluedness instead of non locality}
We can immediately see some similar features between the Linde universe multiplication as described by eq. \rf{lindedoubleaction} and the modified measure theory, with the measure assuming a positive or a negative value, as expressed by eq, \rf{solutions of corrected}, instead of the obvious non locality of the Linde approch, the modified measure approach can offer instead multi valued feature of the $\varphi_a$ space with respect to the  $x^\m$ space. The double solution for the measure  \rf{solutions of corrected} can be valid for the same coordinate  $x^\m$ , which may correspond however to non unique values in  the $\varphi_a$ space. The doubling  of the measure   \rf{solutions of corrected} has its correspondence in the signed reparametrization invariant formulation of modified measure  \ct{stringsandantistrings}  and the corresponding existence of strings and antistrings as well as branes and anti branes in such fromulation.  

\section{Turning Manifolds with Boundaries into Manifolds without Boundaries,  the problem of time in Quantum Cosmology,  Quantum Space Time,
space times and anti space times}
As we will discuss in this section, these concepts can have interesting applications to quantum cosmology: 1) the consideration of one of the measure fields as a time, instead of the ordinary coordinate time can transform a manifold with boundaries, as a universe with an origin of time,  which is difficult to handle, into a manifold which avoids these difficulties in by considering that the fundamental manifold in terms of the measure fields does not have boundaries and 2) the  problem of time, associated with the coordinate time, which produces vanishing Hamiltonians is avoided. Considering instead one of the measure fields as time avoids this problem and lead us to a new formulation of a quantum space time, 3) using the measure fields as the basis to construct the quantum space time 4) defining space times and antispacetimes, this is in fact related to 1).

Indeed, concerning the first point, in most cosmological models we assume that the universe has an origin in time, this leads then to a manifold with boundaries if we think the fundamental manifold is coordinate space and time. A well known example of such a phenomena is the pair creation in QED, which we review in the next subsection.
 \subsection{ Pair Creation in a Strong Uniform Electric Field in QED, as an example of turning a manifold with boundary in time to a manifold without boundary in proper time}
 From the 'Feynman' perspective, negative energy waves propagating into the past are physically realized as the antiparticles propagating into the future.  As he has shown \cite{Fyn:e-m} that from the classical equations of motion for a particle in an external field can be written as
\begin{equation} \label{LorentzForce1}
m\frac{d^{2}z^{\mu}}{d\tau^{2}}=e\frac{dz_{\nu}}{d\tau} F^{\mu \nu}
\end{equation} 
where $\tau$ is the proper time.  If we note as Feynman has that if we allow $\tau\rightarrow -\tau$ the equation becomes
\begin{equation}
m\frac{d^{2}z^{\mu}}{d\tau^{2}}=-e\frac{dz_{\nu}}{d\tau} F^{\mu \nu}
\end{equation} 
which is identical to the previous equation except that the particles charge has changed.  In other words, as far as its charge is concerned, it has become the antiparticle.  Thus, proper time running backward (i.e., $\tau \rightarrow -\tau $), while keeping the coordinate time unchanged.  led to the particle becoming an antiparticle.
Of course we can take the equivalent, but more suitable for our purposes transformation that we change the direction of coordinate time, while requiring that the proper time remains unchanged.

This would be exactly analogous to taking one of the four scalars as an analogous of the proper time and the coordinate time, which would be another entity, 

We are now in a position where we can discuss the scattering of a  particle in an external field.  Four possibilities are seen to exist: i) the scattering of the particle by the field, ii) the creation of a particle-antiparticle pair.  iii) the annihilation particle-antiparticle pair iv) the scattering of an antiparticle by the field. Below, we review one of them and in the following section we draw conclusions concerning the dynamical string theory and the analogy of the modified measure in the string case with the proper time in the particle case.
 We shall present here a simplified version, by A.Vilenkin \cite{Vilenkin}, of the calculation of pair production first performed by Schwinger \cite{Schwinger1} 
 To begin our discussion, we consider a particle with charge $e$  and mass $m$ in a constant electric field.  The general equation of a particle in a field is most convenient written in terms of the Maxwell tensor $F_{\mu \nu}$ where for a constant electric $\mathbf{E}$ in the $x-$direction, $F^{01}=~-~E, \ \ F^{10}=E \ \ F^{0}_{\ 1}=E, \ \ F^{1}_{\ \ 0}=E$.  More explicitly, we have
 \begin{eqnarray}
 F^{\mu}_{\ \ \nu}=
 \left[ \begin{array}{llll} 
 0 & E & 0 &0 \\
 E & 0 & 0 & 0 \\
 0 & 0 & 0 & 0 \\
 0 & 0 & 0 & 0 
 \end{array} \right]
 \end{eqnarray}
The equation of motion for this  particle in a constant electric field is
\begin{equation}
m\frac{d^{2}x^{\mu}}{d\tau^{2}}=e F^{\mu}_{\ \ \nu}\frac{dx^{\nu}}{d\tau}
\end{equation}
\linebreak
The formal solution of for $u^{\mu}=\frac{dx^{\mu}}{d\tau}$ is $u^{\mu}(\tau)=\exp[\frac{e}{m}F^{\alpha}_{\ \ \beta}\tau]^{\mu}_{\nu}u^{\nu}(0)$
The exponential can be expanded and we have
\begin{eqnarray} \label{expo}
\exp[\frac{e}{m}F^{\alpha}_{\ \ \beta}\tau]^{\mu}_{\nu}=\delta^{\mu}_{\nu}+\frac{e}{m}\tau E \Delta^{\mu}_{\nu}+\frac{1}{2}(\frac{e}{m}\tau E \Delta^{\mu}_{\nu})^{2}+\cdots
\end{eqnarray}
where
\begin{eqnarray}
\Delta=
\left[\begin{array}{llll}
0 & 1 & 0 & 0 \\
1 & 0 & 0 & 0 \\
0 & 0 & 0 & 0 \\
0 & 0 & 0 & 0 
\end{array} \right]
\end{eqnarray}
Separating even and odd power in Eq. \ref{expo} we have
\begin{eqnarray}
u^{0}&=&\cosh\left(\frac{eE\tau}{m}\right)u^{0}(0)+\sinh\left(\frac{eE\tau}{m}\right)u^{1}(0)  \nonumber \\
u^{1}&=&\sinh\left(\frac{eE\tau}{m}\right)u^{0}(0)+\cosh\left(\frac{eE\tau}{m}\right)u^{1}(0)
\end{eqnarray}

$\\$
Integrating with respect to $\tau$ yields (where we have dropped arbitrary constants of integration)
\begin{eqnarray}
x^{0}&=&\frac{m}{eE}\left\{\sinh\left(\frac{eE\tau}{m}\right)u^{0}(0)+\cosh\left(\frac{eE\tau}{m}\right)u^{1}(0)\right\} \nonumber \\
x^{1}&=&\frac{m}{eE}\left\{\cosh\left(\frac{eE\tau}{m}\right)u^{0}(0)+\sinh\left(\frac{eE\tau}{m}\right)u^{1}(0)\right\}
\end{eqnarray}

We choose the following boundary conditions $u^{1}(0)=0,\ \ u^{0}(0)=1$ which leads to
\begin{eqnarray} \label{particle}
x^{0}&=&\frac{m}{eE}\sinh\left(\frac{eE\tau}{m}\right) \nonumber \\
x^{1}&=&\frac{m}{eE}\cosh\left(\frac{eE\tau}{m}\right)
\end{eqnarray}
and for the boundary condition  $u^{1}(0)=0,\ \ u^{0}(0)=-1$  leads to
\begin{eqnarray} \label{antiparticle}
x^{0}&=&-\frac{m}{eE}\sinh\left(\frac{eE\tau}{m}\right) \nonumber \\
x^{1}&=&-\frac{m}{eE}\cosh\left(\frac{eE\tau}{m}\right)
\end{eqnarray}
The solution given by Eq. \ref{particle} represents a particle solution while the solution of Eq. \ref{antiparticle} represents the anti-particle solution. Both solutions, together satisfy

\begin{equation} \label{hyperbola} (x^{1})^{2}-(x^{0})^{2}=\left(\frac{m}{eE}\right)^{2}\end{equation}

At classical level, these solutions are distinct and one solution can not evolved into the other.  Thus, a particle couldn't evolve to an anti-particle.  However, the semi-classical approximation which consists of considering the classical equations of motion but with imaginary time.  Then inserting $t=-it_{E}$
we obtain that the hyperbola of Eq. \ref{hyperbola} becomes a circle
\begin{equation}
(x^{1})^{2}+t_{E}^{2}=\left(\frac{m}{eE}\right)^{2}
\end{equation}
This tunneling solution can now interpolate between the anti-particle and particle solutions.  
In the imaginary time region, the action, $S=-iS_{E}$ where $S_{E}$ is given by 
\begin{equation}
S_{E}=\int dt_{E} \left\{ m\sqrt{1+\left(\frac{dx}{dt_{E}}\right)^{2}}-eEx\right\}
\end{equation}

Introduction the angular variable $\theta$ where $x=\frac{m}{eE}\cos \theta,\ \ t_{E}=\frac{m}{eE}\sin \theta$ we obtain that $S_{E}=\pi m^{2}/eE$.  Since the probability is given (up to pre factors) by $\exp{(-S_{E})}=\exp(-\frac{\pi m^{2}}{eE})$ for $eE>0$.
Notice that the distance of the particle and antiparticle at the moment of creation is $\Delta x=\frac{2m}{eE}$ which has a physical interpretation is manifest in writing it as 

\begin{equation}
W=eE\Delta x=2mc^{2}
\end{equation}
$\\$
where we have restored the 'c' to make its physical meaning clearer.  Thus we can create a pair of particle-antiparticle in a constant electric field by performing work, $W$ in a distance $\Delta x$ equal to the the sum of the rest masses of the particles, i.e. $2mc^{2}$. In the case for an electric field that doesn't extend through all of space, still it must be extended enough to perform the work equal to the sum rest masses of the two particles in order to create a pair.

From the point of view of our discussion here, we see that in the pair creation process, the proper time proceeds from minus infinity to plus infinity, while for this manifold the minimum coordinate time sits in euclidean space at   $t_{E}=-\frac{m}{eE}$, the coordinate manifold is a manifold with boundaries, while the proper time manifold does not have boundaries associated with it.

\subsection{Resolution of The problem of Time in Quantum Cosmology using one of the measure fields as time}
A related difficulty related to the use of coordinate time in quantum cosmology is that the invariance under reparametrizations in the coordinate time leads to the constraint that the Hamiltonian equals zero. In this case the analogous of rge Schroedinger equation  for Quantum gravity will tell us that the wave function is time independent. 

In the presence of this problem, the ¨time problem in Quantum Cosmology¨, many proposals for alternative fields to replace the cosmic time were proposed, for a review see \cite{PROBLEmoftimeinQC}.

To choose the corresponding time among the four measure fields, we can use the volume preserving diffeomorphisms \rf{VOLPRESDIFF} or the general coordinate invariance to choose three 
of the measure scalars as the three spacial coordinates $x, y, z$, the remaining scalar can be defined as the time like coordinate, which cas run in the same direction of the coordinate time or against.
\subsection{Quantum Space Time}
By considering a gauge (of the volume preserving diffeomorphisms or the general coordinate  invariance groups )  where three of the coordinates $(t, x, y, z)$, say $ (x,y,z)$ are set to be equal to three of the measure fields, we define a procedure by means of which coordinates become dynamical , since the measure fields are real dynamical variables and space is now  represented by the measure fields. Notice that these
measure fields have non vanishing canonically  conjugated momenta, etc. so they represent the dynamical and quantized  space time (for some papers and a  review of quantum space time see \cite{quantumspacetime}). 

One can see that the for the canonically conjugate momenta for the measure fields, the system shows constraints that relate the canonically conjugate momenta with the spacial derivatives of these measure gauge fields,
indeed,
\begin{equation}
\pi^ {\varphi_a} = M \epsilon^{a a_1 a_2 a_3}  \epsilon^{ijk}\pa_i\varphi_{a_1} \pa_j\varphi_{a_2}\pa_k\varphi_{a_3} 
\end{equation}
Showing a constraint that relates the  canonically conjugate momenta of the measure fields with the spacial derivatives of these fields. The commutators could be calculated not from the Poison brackets but from the Dirac brackets \cite{DIRACBRACKETS}, that are  defined from the Poisson brackets of  the constraints, including the gauge fixing, like the ones we have mentioned. On the mass shell $M$ is a constant, but off the mass shell it must be defined as $ M = R + L  +2 \Omega \frac{\P (H)}{{(-g)}} $. The quadratic term in $\Omega$ in the action implies that the dependence of the action on the measure time derivatives is not linear, which is good because lagrangian systems linear in the time derivatives of some variable have some difficulties.

If the Dirac brackets between the measure fields can be non vanishing, even if the Poisson brackets are zero, we could obtain a non commutative space time .
\subsection{Particles vs antiparticles, Strings vs anti Stings , Branes vs anti Branes and  Space Time vs anti Space Time}
We have reviewed already the discussion of particles vs antiparticles due to Feynman, which consisted of interpreting antiparticles as particles where the proper time moves in the opposite direction of the coordinate time. 

In the case of extended objects, which also can be formulated in terms of modified measures, as we have mentioned before, and in particular, one can advance the concepts of strings vs anti strings and that of branes vs anti branes
 \cite{stringsandantistrings} . The antistrings are identified, in analogy with the Feynman picture for antiparticles when the sign of the Riemannian measure is opposite to the metric independent measure, defined in terms of world sheet scalars, so the same rule should be applied with the identification of the anti space time in comparison with the space time.
 To see this more clearly, in the case of a string, we can use the volume preserving diffeomorphisms to set one measure field to a spatial world sheet variable, then the remaining measure field is a time like coordinate that can go in the direction of the world sheet coordinate or the opposite, which is the difference between strings and anti strings and similar for branes and anti branes.

 For a $4D$ space time (or we can say a space time filling brane), we use the volume preserving diffeomorphisms to set a gauge where three measure fields are set equal to three spatial coordinates, then the remaining measure field is the analog of the proper time in the particle case, which can run in the same direction or oposite to the coordinate time, defining a space time or an anti space time. 
\section{Discussion}
We have discussed how general coordinate invariance is extended to general coordinate transformations that have a negative jacobian, one way to achieve this invariance is in the case the coordinate has a negative jacobian in all spacetime. 


In Linde´s scenario,  as long as the jacobian is uniformly negative over all space, invariance will be achieved if the same transformation is performed both in the $x$ and $y$ coordinates. No modifications of integration manifolds are required in this case,

A local realization of signed general coordinate transformation invariance is possible by introducing a non Riemannian Measure of integration, which transforms according to the jacobian of the coordinate transformation, not the absolute value of the  jacobian of the coordinate transformation as it  is the case with $\sqrt{-g}$. The theory studied here could provide a way to achieve similar similar effects to that of Linde´s universe multiplication scenario, but without non localities. 

This analysis may also be applied to give a framework for the Farhi, Guven and Guth treatment of the Quantum creation of a baby Universe, which just use Einstein´s equations, as we have also, for a false vacuum bubble and find that the self consistent solution must have regions where the measure of integration becomes negative, although GR itself does not provide for a well defined mechanism to produce negative measures of integration.

This negative measure issue is manifested in a somewhat hidden way in the analysis of the second reference in \cite{FischlerPolchinski},  the problem is that (see discussion
around eqn. (59) in this paper) the Euclidean lapse changes
sign within the integration range. Notice that only the square of the lapse function has meaning in GR and the lapse function itself would only appear in the square root of the determinant of the determinant of the metric, assigning an arbitrary  sign to the lapse function corresponds to an arbitrary prescription for the definition of the measure of integration, something we avoid here by stating that the measure is an independent object independent of the metric.

The modified measure theory we have described here does this and the sign in \rf{solutions of corrected} will  be dynamically determined, matching the solutions of Farhi, Guven and Guth. The Farhi, Guven and Guth also require an additional space which is multivalued with respect to the $x$ space and regions with negative measure  to be able to describe the tunneling solutions and the modified measure and a non trivial mapping from the measure fields to the $x$ space can serve this purpose.

The Linde universe multiplication, with one universe associated with a negative measure is a simpler application of these ideas, since it can be formulated already at the classical level. Again we would rely on multi valuedness of the mapping of the coordinate space to the 4 scalars measure space and of course the possibility of a positive and a negative measure for space time.

One could also go beyond General Relativity like theories, and generalize the modified measure theory \rf{TMMT1} in the following way, 
\be
S = \int d^4 x\,\P_1 (A) \Bigl\lb R + L^{(1)} \Bigr\rb +  
\int d^4 x\,\P_2 (B) \Bigl\lb L^{(2)} + \eps R^2 + 
\P_2 (B)\frac{\P (H)}{(-g)}\Bigr\rb \; .
\lab{TMMTsigned}
\ee
this will be now also a signed general coordinate invariant theory. One can again explore a representation of modified measures in terms of 4 scalar fields, etc. This case will be more involved, requiring solving for two measures, defining an Einstein frame, etc, following for example the work done in \cite{quintess}. This will be done in a future publication.

Going back to what we have already done in this paper, in the  negative  $\Omega$   measure case, as defined in \rf{omega}, this represents a different orientation of the $\varphi_a$ space with respect to the $x^\m$ space, it is reminiscent to the  interpretation by Feynman of antiparticles as negative energy particles traveling backwards in time, which he also described as a situation where laboratory time and the proper time of the particle run in different direction, so one could think of  $\Omega$  negative as anti space. In this context the Farhi, Guth and Guven \cite{FARHIGUVEN} , which is a creation of a baby universe that seems to be associated to  pair creation of spaces and anti spaces. It is interesting to note the analogous observations in Linde´s universe multiplication scenario \cite{Lindemultiplication} concerning the particle and antiparticles.

The non locality in the universe multiplication scenario shows in the equations of motion in  the form of space time averages of field variables over the 
complete history of the Universe which appear in the equations of motion as effective coupling constants. These integrals could be infrared divergences, depending on the assumed history of the universe \cite{Lindemultiplication}.
And assuming complete symmetry between the negative measure universe and the positive measure it is possible to argue for a zero effective cosmological constant .
The modified measures approach in contrast is local , however if the set of four scalars that define the measure is introduced, and the mapping of the coordinates to the four scalars is multi valued, this could give rise then to the appearance of non locality , although there would be no reason for complete symmetry between positive and negative measure regions and therefore unlike in Linde´s scenario, for an exact cancellation of an exactly zero effective cosmological constant. Alternatively, to get similar effects to the Linde model with our local theory, one could assume entanglement between positive and negative measure regions of space,

Finally, we have discussed applications of these concepts to quantum cosmology: 1) the consideration of one of the measure fields as a time, instead of the ordinary coordinate time can transform a manifold with boundaries, which is difficult to handle into a manifold which avoids these difficulties, when this happens, we also see that the correspondance between measure fields and standard coordinates is multivalued, just as in the case of particle antiparticle creation in QED, where there are no boundaries for the proper time, but the coorniate time has a minimum value,  2) the  problem of time, associated with the coordinate time, which produces vanishing Hamiltonians is avoided, 3) a possible new avenue to the construction of quantum space time is proposed, 4) the identification of space times vs anti space times, that can play an important role in quantum cosmology. The Linde model appears as a non local coexistence of spaces and antispaces
and manifolds containing spaces and anti spaces can easily avoid boundaries that would appear in the naive coordinate space as we mentioned before.


\begin{acknowledgements}
I want to thank Stefano Ansoldi,  Alex Kaganovich , Emil Nissimov and Svetlana Pacheva   for conversations,  FQXi  for great financial support for work on this project at BASIC in Ocean Heights, Stella Maris, Long Island,  Bahamas, and CosmoVerse • COST Action CA21136 and  very special thanks to CA18108 - Quantum gravity phenomenology in the multi-messenger approach for financial support and for invitation to the COST Action: CA18108 - Quantum gravity phenomenology in the multi-messenger approach Final Annual Conference where this paper was presented.
\end{acknowledgements}

\end{document}